\documentclass[fleqn,12pt,twoside]{article}
\usepackage[headings]{espcrc1}

\readRCS
$Id: espcrc1.tex,v 1.2 2004/02/24 11:22:11 spepping Exp $
\usepackage{graphicx}
\usepackage[figuresright]{rotating}
\usepackage[figuresright]{rotating}

\newcommand{\AmS}{{\protect\the\textfont2
  A\kern-.1667em\lower.5ex\hbox{M}\kern-.125emS}}

\title{The azimuthal anisotropy of electrons from heavy flavor decays in $\sqrt{s_{NN}}$ = 200 GeV Au-Au collisions by PHENIX}

\author{Shingo Sakai \address{Institute of Physics, University of Tsukuba, Tsukuba, Ibaraki 305, Japan} for the PHENIX Collaboration\thanks{for the full list of PHENIX authors and acknowledgements, see Appendix 'Collaboration' of this volume.}}

\begin{document}

\maketitle

\begin{abstract}
The transverse momentum dependence of the azimuthal anisotropy parameter $v_{2}$,
the second harmonic of the azimuthal distribution, for electrons 
at mid-rapidity ($|\eta|<0.35$) has been measured with the PHENIX detector 
in Au+Au collisions at $\sqrt{s_{NN}}$ = 200 GeV with respect to the reaction plane 
defined at high rapidities ($|\eta|=3-4$).
From the result we have calculated non-photonic electron $v_{2}$, which is expected to reflect 
the heavy-flavor azimuthal anisotropy,
by subtracting $v_{2}$ of electrons from other sources such as photon conversions,
Dalitz decay etc.
 
\end{abstract}

\section{Introduction}
The azimuthal anisotropy of particle emissions is a powerful tool to 
study the early stage of ultra-relativistic nuclear collisions. 
Previous measurements of $v_{2}$ for light hadrons,
such as pions and kaons~\cite{pidflowPHENIX}, are consistent with the quark coalescence model~\cite{qcoales},  
which assumes that the $v_{2}$ of light hadrons comes from their parent quarks.
This suggests that the $v_{2}$ already develops in the 
partonic phase for hadrons made of light quarks. 
Heavy flavor $v_{2}$ measurement will provide important information on the
origin of the flow due to the much larger mass.

Heavy quark production is  well studied 
by measuring electrons from their semi-leptonic decays 
in the PHENIX experiment at RHIC ~\cite{esingle130}~\cite{esingle200}. 
We have studied the azimuthal anisotropy of heavy flavor 
by measuring the $v_{2}$ of electrons from semi-leptonic heavy flavor decays 
in Au+Au collisions at $\sqrt{s_{NN}}$ = 200 GeV.
 
\section{Analysis}
The inclusive electron sample has two components: (1)``non-photonic'' - primarily semi-leptonic decays of mesons
containing heavy (charm and bottom) quarks, and (2)``photonic'' - Dalitz decays of light neutral mesons 
($\pi_{0}$, $\eta$, $\eta^{\prime}$, $\omega$ and $\phi$)
and photon conversions in the detector material ~\cite{esingle200}.
The azimuthal distribution of electrons ($dN^{e}/d\phi$) is the sum of the azimuthal distributions of photonic 
electrons ($dN^{\gamma}/d\phi$) and non-photonic electrons ($dN^{non-\gamma}/d\phi$):
\begin{eqnarray}
\frac{dN_{e}}{d\phi}&=&\frac{dN_{e}^{\gamma}}{d\phi}+\frac{dN_{e}^{non-\gamma}}{d\phi}.
\label{eq:EQ1}
\end{eqnarray}
From Eq. 1 the non-photonic electron $v_{2}$ ($v_{2}^{non-\gamma}$) is expressed as
\begin{equation}
v_{2_{e}}^{non-\gamma} = \frac{(1+R_{NP})v_{2_{e}}-v_{2_{e}}^{\gamma}}{R_{NP}}.
\end{equation}
where $v_{2_{e}}$ is the inclusive electron $v_{2}$, $v_{2_{e}}^{\gamma}$ is the photonic electron $v_{2}$ 
and $R_{NP}$ is the ratio of the number of non-photonic electrons to photonic electrons ($N_{e}^{non-\gamma}/N_{e}^{\gamma}$).
We experimentally determined the ratio from analysis of special runs in which a photon converter was installed.
The details of the method are described in \cite{esingle200}.

The transverse momentum dependence of the inclusive electron $v_{2}$ measured by the reaction plane method ~\cite{rpmethod} is shown in Fig.~\ref{fig:Ev2}.
The reaction plane is determined by  the beam-beam counters and 
the electrons are identified by the ring imaging Cherenkov counter and the electromagnetic calorimeter.
The photonic electron $v_{2}$ is obtained by two methods:
(1) directly from data with and without the converter and 
(2) Monte Carlo simulation~\cite{ev2_RUN2}.
In the first method, the yield of electron with ($N_{e}^{conv-in}$) and without the converter ($N_{e}^{conv-out}$) can be written as,
\begin{eqnarray}
N_{e}^{conv-in} &=& R_{\gamma}N_{e}^{\gamma} + N_{e}^{non-\gamma} \nonumber \\
N_{e}^{conv-out} &=& N_{e}^{\gamma} + N_{e}^{non-\gamma}.
\end{eqnarray}
where $R_{\gamma}$ is the ratio of the number of photonic electrons with and without the converter.
From Eq. 3 the relation between the number of electrons and the values of $v_{2}$ are given as:
\begin{eqnarray}
N_{e}^{conv-in} v_{2}^{conv-in}    &=& R_{\gamma}N_{e}^{\gamma} v_{2_{e}}^{\gamma} + N_{e}^{non-\gamma} v_{2_{e}}^{non-\gamma} \nonumber \\
N_{e}^{conv-out} v_{2}^{conv-out} &=& N_{e}^{\gamma} v_{2}^{\gamma}+ N_{e}^{non-\gamma} v_{2}^{non-\gamma}.
\end{eqnarray}
where $v_{2_{e}}^{conv-in}$ is inclusive electron $v_{2}$
measured with the converter and $v_{2_{e}}^{conv-out}$ is inclusive electron $v_{2}$ measured without the converter.
From the Eq. 4 the photonic electron $v_{2}$ is given as:
\begin{equation}
v_{2_{e}}^{\gamma} = \frac{ (1+R_{NP})v_{2_{e}}^{conv-out} - (R_{\gamma}+R_{NP})v_{2_{e}}^{conv-in}}{(1-R_{\gamma})}.
\end{equation}
The photonic electron $v_{2}$ determined from the first method (open circles) 
and from the second method (solid line) are shown in Fig.~\ref{fig:Ev2}.
Due to the limited statistics of the runs with the converter,
the second method is applied above 1.0 GeV/$c$ in this analysis.
From the result the inclusive electron $v_{2}$ is smaller than the photonic electron $v_{2}$, which means that the 
non-photonic electron $v_{2}$ is smaller than both of them.
The transverse momentum dependence of $R_{NP}$ is shown in Fig.~\ref{fig:Rnp}.
Above 1.0 GeV/$c$ more than 50 $\%$ of electrons come from the non-photonic component.

\begin{figure}[tb]
\begin{minipage}[t]{73mm}
\includegraphics[width=21pc, height=17pc]{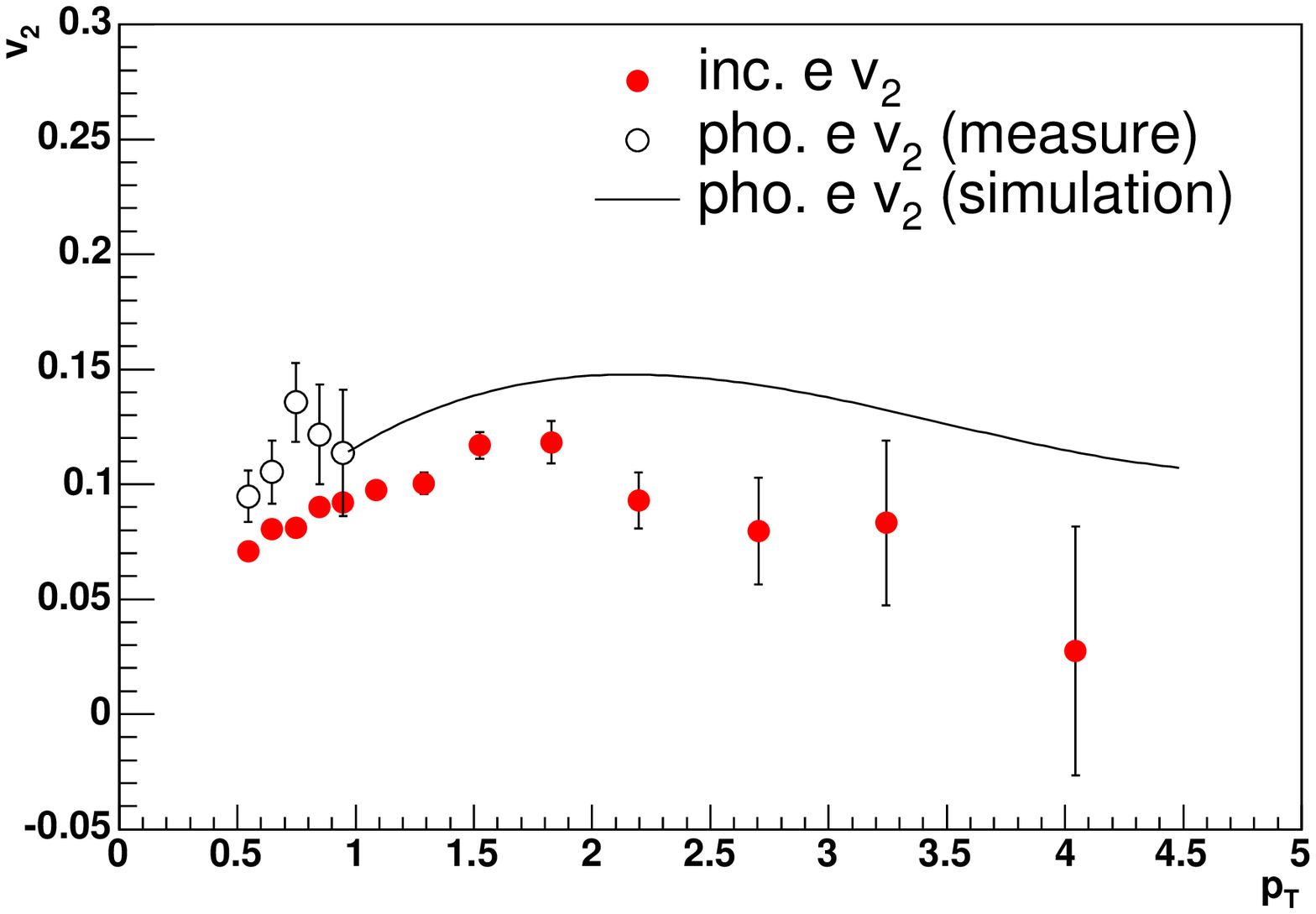}
\caption{\label{fig:Ev2} The transverse momentum dependence of the inclusive electron $v_{2}$ (closed circle) 
and the photonic electron $v_{2}$ (open circle and solid line).}
\label{fig:Ev2}
\end{minipage}
\hspace{\fill}
\begin{minipage}[t]{73mm}
\includegraphics[width=21pc, height=17pc]{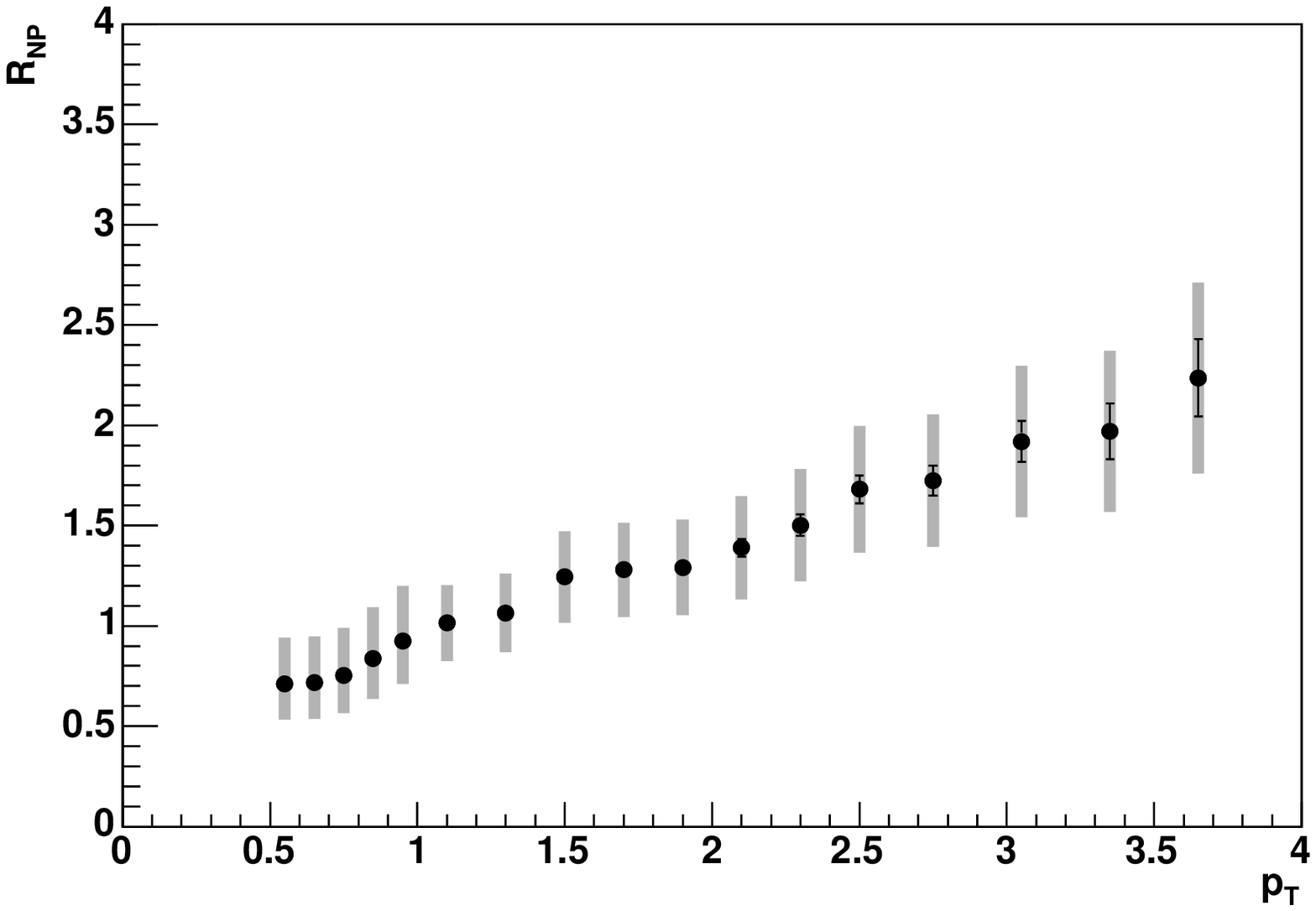}
\caption{\label{fig:Rnp} The transverse momentum dependence of the ratio 
of the number of non-photonic electrons to photonic electrons ($N_{e}^{non-\gamma}/N_{e}^{\gamma}$).} 
\label{fig:Rnp}
\end{minipage}

\end{figure}

\section{Results}
The transverse momentum dependence of the non-photonic electron $v_{2}$ is shown in the Fig.~\ref{fig:non_phov2_0}.
The statistical errors are shown as vertical lines and the 1 $\sigma$ systematic uncertainties are shown as brackets 
in the figure. Assuming the quark coalescence model,
decay electron $v_{2}$ from $D$ mesons has been predicted~\cite{rapD}.
In the model $D$ mesons are formed from charm quark coalescence with thermal light quarks at hadronization. 
For charm quark momentum spectra, two extreme scenarios are considered. 
The first scenario assumes no reinteractions after the production of charm-anticharm quark 
pairs in initial state hard processes (calculated from PYTHIA). 
The second scenario assumes complete thermalization with the transverse flow of the bulk matter.
The heavy flavor electron $v_{2}$ with decay electrons from $D$ mesons in the 
``no reinteraction'' scenario is shown as a solid line, while the ``thermalization'' scenario
is shown as a dashed line in the figure.
Below 2.0 GeV/$c$ the non-photonic electron $v_{2}$ is in good agreement with the electron $v_{2}$ obtained by assuming 
charm quark flow.

\begin{figure}[tb]
\begin{minipage}[t]{73mm}
\includegraphics[width=21pc, height=17pc]{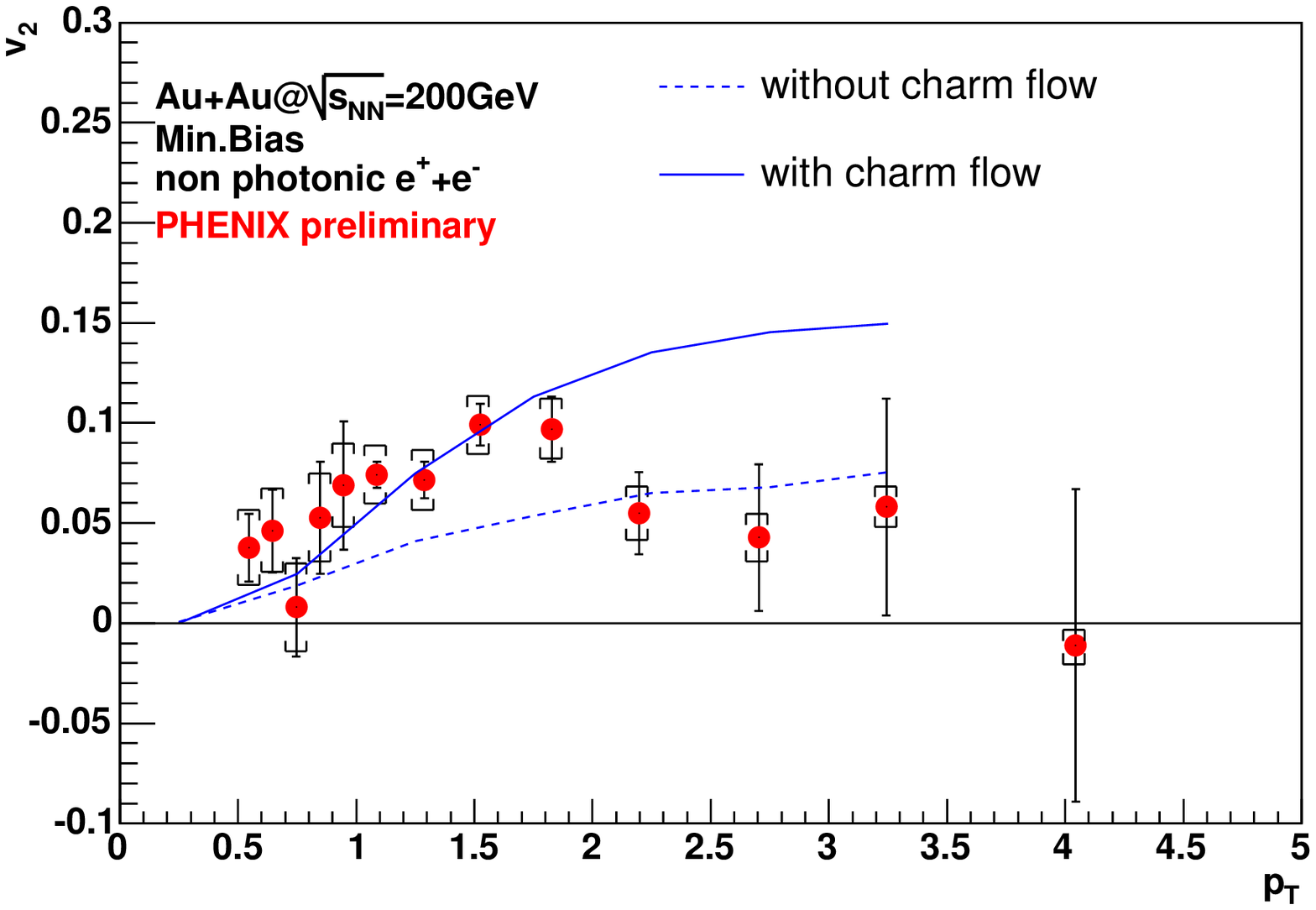}
\caption{\label{fig:non_phov2_0} The transverse momentum dependence of the non-photonic electron $v_{2}$ and 
the model prediction with and without charm flow \cite{rapD}.} 
\label{fig:non_phov2_0}
\end{minipage}
\hspace{\fill}
\begin{minipage}[t]{73mm}
\includegraphics[width=21pc, height=17pc]{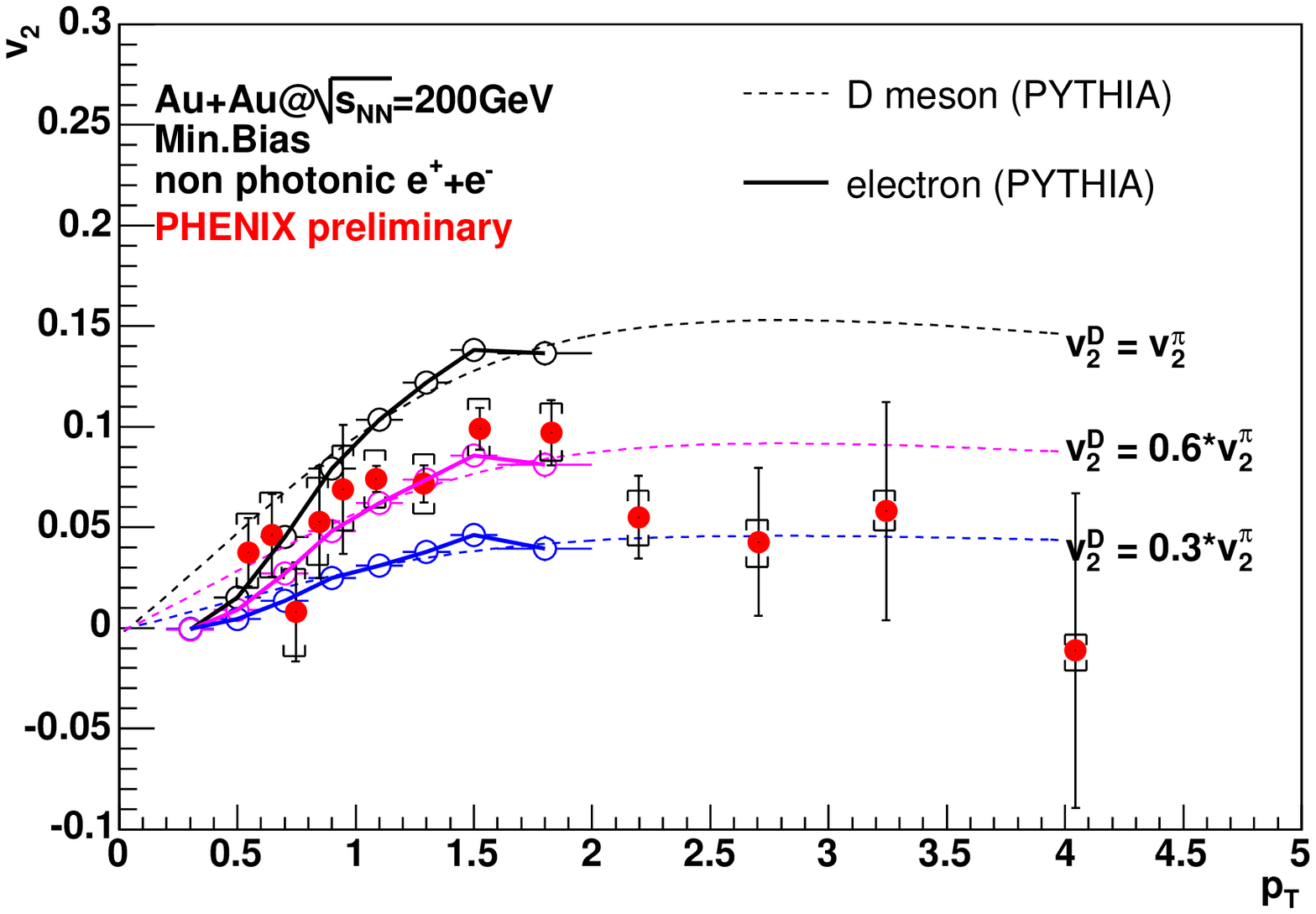}
\caption{\label{fig:alpha_pt2} Electron $v_{2}$ from $D$ meson assuming the shape is proportional to pion $v_{2}$}
\label{fig:non_phov2_1}
\end{minipage}
\end{figure}

We also compare the non-photonic electron $v_{2}$ with the simulation of electron $v_{2}$ from $D$ meson 
assuming the shape is proportional to pion $v_{2}$ \cite{pidflowPHENIX} as  
\begin{equation}
v_{2}^{D}(p_{T})= a v_{2}^{\pi}(p_{T})
\end{equation}
here $a$ is a scale factor of the pion $v_{2}$.
We calculate the $v_{2}$ with several $a$ parameter (30 $\%$, 60 $\%$ and 100 $\%$). 
The comparison of the measured non-photonic electron $v_{2}$ and the simulation results are shown in the Fig.~\ref{fig:non_phov2_1}.
The solid line is $D$ meson $v_{2}$ and the open circle plot is decay electron  $v_{2}$.
The non-photonic electron $v_{2}$ seems to prefer the $D$ meson $v_{2}$ assuming 60 $\%$ of the pion $v_{2}$.
It might be suggested that $D$ meson has non-zero $v_{2}$ and has smaller $v_{2}$ than the pion.

\section{Summary}
In this paper we present the methods of measuring non-photonic electron
v2 and show the results.
We compare our results with two extreme model predictions which assume no charm quark flow or complete thermalization 
of charm quarks in the medium.
Below 2.0 GeV/$c$ the non-photonic electron $v_{2}$ is in good agreement with the electron $v_{2}$ obtained by assuming 
the complete thermalization of charm quarks.
In addition, we calculate electron $v_{2}$ from $D$ meson assuming the shape is proportional to pion $v_{2}$.
The non-photonic electron $v_{2}$ seems to prefer the $D$ meson $v_{2}$ assuming 60$\%$ of pion $v_{2}$.
It might be suggested that $D$ meson has non-zero $v_{2}$ but smaller $v_{2}$ than that of pions.

\end{document}